# The effect of "Open Access" upon citation impact: An analysis of ArXiv's Condensed Matter Section

**Henk F. Moed**

Centre for Science and Technology Studies (CWTS)
Leiden University, PO Box 9555, 2300 RB Leiden, the Netherlands
Email: moed@cwts.leidenuniv.nl
Tel ++31 71 527 3940
Fax ++31 71 527 3911

## Abstract

This article statistically analyses how the citation impact of articles deposited in the Condensed Matter section of the preprint server ArXiv (hosted by Cornell University), and subsequently published in a scientific journal, compares to that of articles in the same journal that were *not* deposited in that archive. Its principal aim is to further illustrate and roughly estimate the effect of two factors, 'early view' and 'quality bias', upon differences in citation impact between these two sets of papers, using citation data from Thomson Scientific's Web of Science. It presents estimates for a number of journals in the field of condensed matter physics. In order to discriminate between an 'open access' effect and an early view effect, longitudinal citation data was analysed covering a time period as long as 7 years. Quality bias was measured by calculating ArXiv citation impact differentials at the level of individual authors publishing in a journal, taking into account co-authorship. The analysis provided evidence of a strong quality bias and early view effect. Correcting for these effects, there is in a sample of 6 condensed matter physics journals studied in detail, *no* sign of a general 'open access advantage' of papers deposited in ArXiv. The study does provide evidence that ArXiv *accelerates* citation, due to the fact that that ArXiv makes papers *earlier* available rather than that it makes papers *freely* available.

## 1    Introduction

The debate on costs and benefits of "Open Access" compared to other forms of scientific literature publishing has a political, an economical and an information–scientific dimension. In this debate, the term "Open Access" has different meanings. It is used to indicate a particular business model of scientific publishing, in which essentially the authors of articles published in a journal pay the costs of the publication, and their full texts are freely accessible once they are published. But the term "Open Access" is also used to indicate open or free accessibility of scientific documents in general, regardless of whether these are published in a journal running under an Open Access model, or published in a journal applying other business models but also (often after several months) deposited in a freely accessible archive such as a personal website or an institutional depository, or as pre-prints in a freely accessible pre-print server.

From an information–scientific perspective, the key issue in this debate is how scientific–scholarly communication, and particularly its publication processes, can optimally profit from the new developments in information and communication technologies. From this perspective, it is highly relevant to analyse and evaluate the feasibility of the various publication models and their effects, both at a short and at a longer term. It is no wonder that citation analysis constitutes one of the principal tools in this research. More than three decades ago Eugene Garfield, the founder of the Science Citation Index, showed how citation analysis can be used to study the scientific–scholarly communication system, and to contribute to its better functioning and hence to a better science (Garfield, 1972). He and his followers illustrated this in numerous studies.

During the past years, several case studies applied citation analysis to examine the effects of Open Access business models or openly accessible publication archives upon the 'visibility' or 'impact' of published articles (e.g., Harnad and Brody (2004), Kurtz et al. (2005), Davis and Fromerth (2006), Eysenbach (2006)). These studies explore statistical relationships among variables of interest, in case studies examining particular data samples, variables and access modalities. The study presented in this article is also a case study, primarily of a methodological nature. It relates to papers deposited in the Condensed Matter section of ArXiv, a preprint server founded by Ginsparg, and currently hosted by Cornell University. The key questions this article addresses are:

i. How does the citation impact of articles deposited in ArXiv and subsequently published in a scientific journal compare to that of articles in the same journal that were *not* deposited in that archive?

ii. How should the differences in citation impact among the two sets of articles be explained? Is it only or mainly the open accessibility of ArXiv that accounts for these differences, or are there *other* factors responsible as well, and how strong are their effects?

It builds upon the work by Harnad and Brody (2004). It calculates an ArXiv Citation Impact Differential (CID), a measure that is similar to Harnad and Brody's 'Open Access (OA) to non-OA Impact Ratios', but more appropriate for application at the level of individual authors. Results were compared to those presented by Harnad and Brody. A strict replication of their findings could not be carried out since their data related to other ArXiv sections.

Following the work by Kurtz et al. (2005) and Davis & Fromerth (2006), three effects were distinguished. The first is the genuine *open access effect*, in the sense that ArXiv increases access to research papers. It needs emphasising here that none of the journals analysed in this paper have adopted an Open Access business model. The second effect is termed the *early view effect*: articles appear earlier in ArXiv than they do in the (electronic or printed) journal. The aspect of accessibility at stake here is '*earlier*' versus '*later*', distinct from 'open (or free)' versus 'not open', as in the open access effect. Finally there is a *self-selection* effect or *quality bias*. Kurtz et al. distinguished two dimensions. The first is that prominent authors may tend to deposit their papers in ArXiv more often than less prominent scientists do. In other words, prominent authors may be overrepresented in ArXiv. The second is that authors – be it prominent or less so – may tend to deposit their better papers in ArXiv.

The principal aim of the work described in this paper is to further illustrate and roughly estimate the early view effect and quality bias of ArXiv upon citation impact. It presents estimates for a number of journals in the field of condensed matter physics. In order to discriminate between the open access effect and the early view effect, longitudinal citation data is analysed covering a time period of 7 years, which is much longer than the time period of 18 months considered in a recent paper by Eysenbach (2006) on the "citation advantage" of "OA" papers published in the Proceedings of the National Academy of Sciences.

The structure of the paper is as follows. Section 2 describes data collection and elementary data handling, as well as methodological issues. The empirical results are presented in Section 3. Finally, Section 4 gives a discussion, draws conclusions and makes suggestions for further research.

## 2    Data and methods

A database was created of all 74,521 papers deposited in the Condensed Matter section of ArXiv (denoted as ArXiv-CM) during the time period 1992–2005. ArXiv-CM papers were linked to articles in journals processed by Thomson Scientific for the Web of Science (WoS), on the basis of first author names, significant words from the papers' titles and the information available in the 'journal reference field' of an ArXiv publication record. The latter field, designed to give the source (journal, proceedings, book) in which a final version of the paper was published, was filled in only about 40 per cent of the papers. Therefore, it was necessary to search for matches also on the basis of author names and title words. A base assumption underlying this approach is that documents from the two databases linked in this way represent one and the same article, and that ArXiv papers not found in the WoS were *not* (yet) published in WoS journals.

About 75 per cent of ArXiv-CM papers were linked in this way to a WoS source article. In this set of papers, the median time period between the date a paper was deposited in Arxiv and the date it was published in a journal, was found to be about six months. They were published in several hundreds of journals, revealing a skewed distribution of publications among journals. Three journals – *Physical Review B*, *Physical Review Letters* and *Physical Review E* – accounted for 50 per cent of all linked articles. Of the 68 journals assigned in the WoS to the journal category 'Physics, Condensed Matter', 24 had published at least 10 articles, or more than one per cent of its articles, linked to an ArXiv-CM paper. It is on this set of 24 condensed matter physics journals that the analyses presented below are based. The six journals with the largest number of articles linked to an ArXiv-CM paper are listed in Table 1 in Section 3.

Citations to ArXiv versions of papers were traced in the WoS database by applying a citation match-key that included parts of the first author name and the ArXiv paper number. Citations to WoS articles were collected using an advanced matching algorithm that takes into account numerous variations, errors or discrepancies between cited reference and intended target article (Moed, 2005). Author self citations were *not* included in the citation counts. The study described in this paper did *not* analyse citations *within* the ArXiv, from one ArXiv article to another.

The average citation impact of a journal's papers deposited in ArXiv-CM was compared to that of its articles *not* deposited in that archive. If **CPP** denotes the number of received citations per article, and the lowerscripts *a* and **na** whether the cited paper was deposited in ArXiv or not, respectively, Harnad and Brody (2004) defined their OA versus non-OA Impact Ratio (IR) as **IR**=100*(**CPPa**/**CPPna**). Apart from the fact that this ratio obtains a value of 100 per cent if there is no 'open access advantage' at all (in that case, the numerator and denominator have the same value), the main problem of using this ratio is that it may reach extremely high values if **CPPna** is much smaller than one, and especially that it is undefined if this denominator equals zero.

As long as the ratio is calculated for a journal as a whole or for a set of journals, this is normally not a problem. But this paper analyses impact ratios at the level of individual authors, for which numbers of articles and citations are generally much lower, and a **CPPna** value of zero is no exception. Therefore, in this paper an ArXiv Citation Impact Differential (CID) is calculated, defined as:

**CID**=100*(**CPPa**–**CPPna**) / ((**CPPa**+**CPPna**)/2).

Its values range between –200 (if **CPPa**=0) and +200 (if **CPPna**=0). If both **CPPa** and **CPPna** are zero, its value is defined as 0. CID values are generally lower than those obtained by Harnad and Brody's OA to non-OA Impact Ratio.

In order to analyse the *early view effect*, citation impact and ArXiv Citation Impact Differential (CID) of a set of papers were analysed in relation to their age, defined as the time period between publication date and date of citation. ArXiv CID was calculated for two *fixed* citation time windows: one for the time period involving the first three years after publication (the publication year included), and a second for the fourth until the sixth year of publication date. In some analyses citations were counted on a *monthly* basis as well. Following Harnad and Brody (2004), ArXiv CID was also measured for a *variable* citation time window, starting with the year of publication of an article up until 2005.

A methodological issue of interest is whether citations (given in WoS articles) to ArXiv versions of journal articles should be taken into account. On the one hand, one could argue that it is 'unfair' to include these citations in a comparative analysis of journal papers deposited in ArXiv and articles not deposited in this archive, because the latter do not have earlier versions that can be cited. On the other hand, a base assumption underlying the analysis is that the two versions are different representations of the same paper. Therefore, it was decided to consider and count citations to ArXiv versions as well. When *fixed* citation windows were applied and the time interval between the date of a citation to a paper and the paper's publication date was determined, the publication date of an ArXiv paper receiving a citation was defined as its deposit date in ArXiv.

# 3    Results

## 3.1    *Overall results*

**Figure 1** presents ArXiv Citation Impact Differentials (CID) for the collection of all 24 condensed matter physics journals included in the study, applying a *variable* citation time window.

**Figure 1:  ArXiv CID for 24 journals in condensed matter physics using a variable citation time window**

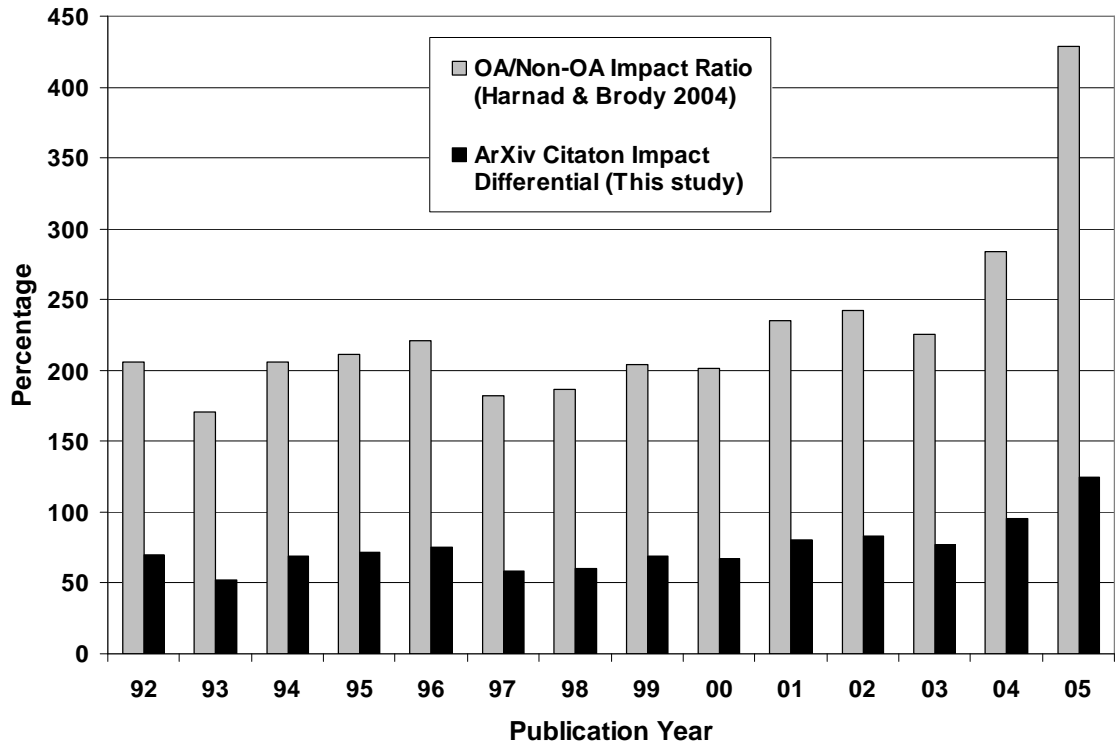

**Legend to Figure 1**: ArXiv CID values on the vertical axis are percentages, not absolute numbers. Citations to ArXiv versions of papers are included in the counts. In this figure, however, the horizontal axis gives the publication year of the journal articles and *not* that of their corresponding ArXiv versions (if there is one). The publication year (deposit date) of the ArXiv versions is on average about half a year earlier. A variable citation window is applied. This means that for instance for papers published in 1992, citations are counted during a 14 year time period 1992-2005, whereas for papers published in 2005 only citations are counted that were received in 2005. In fact, the citation per publication ratio for journal papers deposited in ArXiv gradually decreased from 28.6 for papers published in 1992 to 0.79 for those published in 2005. For journal papers not deposited in ArXiv these ratios are 13.9 and 0.18, respectively.

For the publication years between 1992 and 2003, CID fluctuated between 50 and 75 per cent. In the years 2004 and 2005, it increased substantially and reached values of 96 and 124 per cent, respectively. Figure 1 also shows the values of the 'OA/Non-OA Impact Ratio' as defined by Harnad and Brody (2004). For publication years between 1992 and 2003, this ratio fluctuated between 170 and 225, and reached values of 283 and 428 for papers published in 2004 and 2005, respectively.

## 3.2 Early view effect

The differences in ArXiv CID among publication years in Figure 1 are largely caused by the fact that, when a variable citation time window is applied, the time period over which impact differentials are calculated shortens as the publication year becomes more recent. It is hypothesized that this pattern is mainly due to an early view effect. Figures 2 and 3 further corroborate this hypothesis. They show for articles published during 1996–1999 the number of citations per article on a *monthly* basis during the first seven years after publication date, and compares the age distribution of citations for articles deposited in ArXiv-CM to that of papers that were *not* deposited in that archive.

**Figure 2: Age distribution of citations to papers deposited in ArXiv-CM and to non-deposited papers**

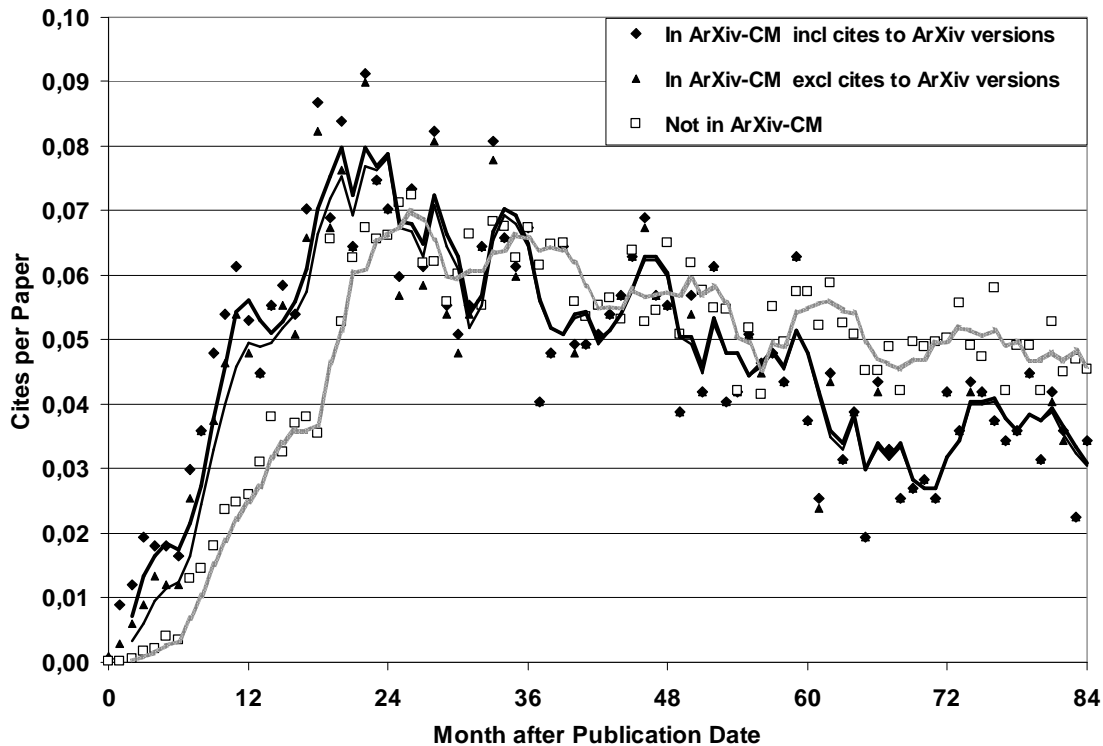

**Legend to Figure 2**. Data relate to articles cited between 3 and 6 times (see main text). The curves represent three months moving averages. Publication date of citing and cited journal articles was measured in this analysis as the date an article was included in the Web of Science. For cites to ArXiv versions of papers the publication date of the cited article is its ArXiv deposit date.

Age distributions of citations to some extent depend upon the number of citations received: high impact papers and poorly cited papers may show different ageing patterns. Therefore, papers were categorized into classes on the basis of their citation frequency, – 1–2, 3–6, 7–18 and >18 citations –, and age distributions were calculated per class. **Figure 2** presents the results for articles cited between 3 and 6 times. Outcomes for the other citation classes were similar.

**Figure 3: Age distributions of citations with the curve for citations to ArXiv-CM papers translated by 6 months**

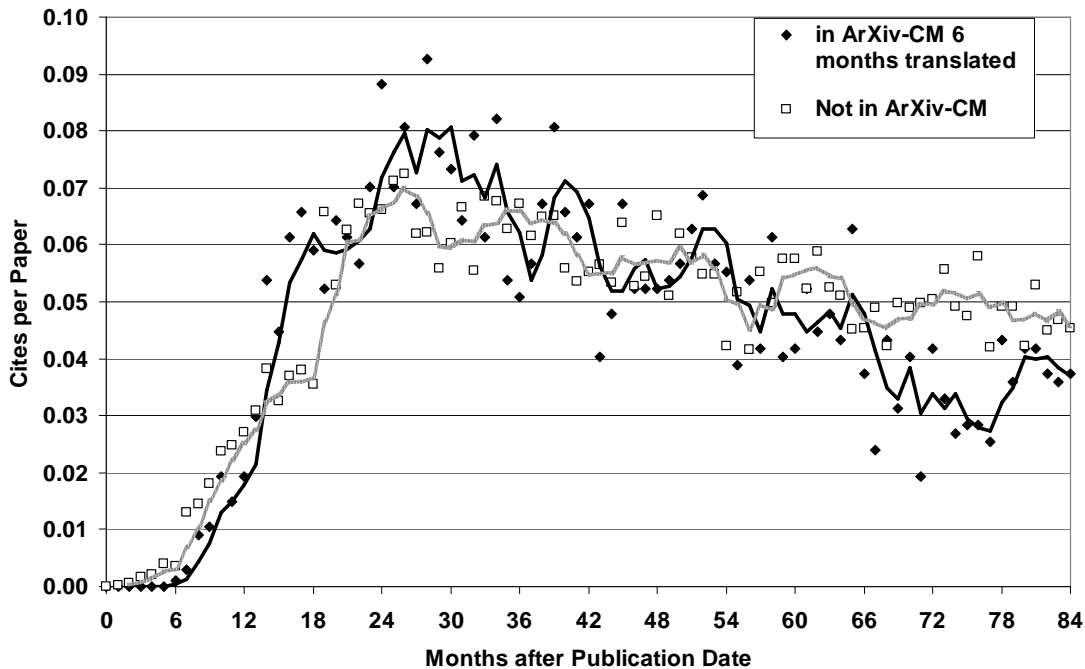

**Legend to Figure 3**. Data relate to articles cited between 3 and 6 times (see main text). The curves represent three months moving averages. Publication date of citing and cited journal articles was measured in this analysis as the date an article was included in the Web of Science. The curve for papers deposited in ArXiv including cites to ArXiv versions in Figure 2 was translated with 6 months to the right and the values of the number of cites per paper during month 1 to 5 was set to zero.

It needs emphasising that the median time interval between the date a paper was deposited in ArXiv and the date it was published in a WoS journal is about 6 months. In **Figure 3** the curve for ArXiv deposited papers shown in Figure 2 is translated with 6 months along the time axis in a positive direction. During the first 24 months this new curve roughly coincides with that for non-deposited papers. Around month 24 both curves reach a maximum, followed by a decline. For papers deposited in ArXiv the maximum is slightly higher and the decline afterward more rapid than it is for articles not deposited in ArXiv.

Figure 3 provides evidence that there is an early view effect at stake. It was therefore decided to apply *fixed* rather than *variable* citation time windows, and calculate ArXiv CID during (a) the first *three* years after publication date (the year of publication included), and (b) during the *fourth to sixth* year after publication. The final row in **Table 1** gives CID for these two fixed citation time windows, calculated for the total collection of 24 journals included in the study. It shows that during the first three years after publication, the ArXiv CID is 80 per cent, and during the fourth to sixth year 64 per cent. The *absolute* decline is 16 per cent (80-64) and the *relative* decline, calculated as ((80-64)/80), amounts to 20 per cent.

**Table 1:    ArXiv Citation Impact Differentials per journal and per citation window**

| Journal | Total Publ 1992-2005 | % Publ in ArXiv | Share of ArXiv publ | *ArXiv Citation Impact Differential* (CID) | |
|---|---|---|---|---|---|
| | | | | *1-3 yrs after publ date* | *4-6 yrs after publ date* |
| Physical Review B | 13,285 | 19.7 % | 70.8 % | 43 % | 27 % |
| European Physical Journal B | 1,195 | 35.4 % | 6.4 % | 87 % | 68 % |
| Journal Physics – Condensed Matter | 1,143 | 7.2 % | 6.1 % | 88 % | 68 % |
| Physica B – Condensed Matter | 523 | 3.0 % | 2.8 % | 83 % | 68 % |
| Solid State Communications | 432 | 4.8 % | 2.3 % | 95 % | 81 % |
| Internat Journal of Modern Physics B | 426 | 8.6 % | 2.2 % | 102 % | 72 % |
| All selected journals (n=24) | 18,757 | 10.2 % | 100.0 % | 80 % | 64 % |

**Legend to Table 1**: Cites to ArXiv versions are included. Table 1 shows that Physical Review B dominates the set of articles published in WoS condensed matter physics journals and deposited in ArXiv-CM, with a share as high as 70.8 per cent. It needs emphasising that the percentage of WoS journals' articles deposited in ArXiv-CM given in Table 1 is an aggregate statistic for the total time period 1992–2005. Calculated on an annual basis, it increased from 0.2 per cent in 1992 to 12.3 per cent in 2000 and to 19.8 per cent in 2005. All journals in Table 1 show a substantial increase over the years.

**Table 1** also presents ArXiv CID for the 6 WoS journals with the largest number of papers deposited in ArXiv-CM. It reveals large differences in CID values among journals, but each journal shows the same pattern as the total collection of 24 journals does: a decline in the CID rate during the fourth to sixth year after publication, compared to that calculated for the first three years after publication date. The mean relative decline rate over these 6 journals is about 24 per cent. *Physical Review B* shows the highest, and *Solid State Communications* the lowest relative decline rate.

### 3.3    *Quality bias*

A first analysis addresses whether prominent authors are overrepresented in the bylines of papers deposited in ArXiv-CM. A methodological problem is how to measure author prominence *independently* of a possible ArXiv advantage effect. Therefore, it was decided to measure author prominence by calculating, on a journal by journal basis, the average number of citations received by an author's papers that were *not* deposited in ArXiv. Two indicators were calculated, one based upon citations received during the first three years after publication date, and a second based upon those received during the fourth to sixth year after publication. Per journal, and for each indicator, authors were categorized into four quartiles containing the top 25 per cent, the 25 per cent above the median but not in the top 25, the 25 per cent below the median but not in the bottom 25 per cent, and the bottom 25 per cent, respectively.

The average number of authors per paper is about 4. There is no one-to-one correspondence between an author and a paper. In this analysis one paper contributes as many times to the counts as the number of authors it has in its byline, so that papers of large author teams have a greater weight than those of small teams. Since co-authorship often reflects collaboration between a senior and one or more junior researchers, this bias can to some extent be reduced by taking into account only 'senior' authors whose publication output exceeds a certain threshold. Therefore, the analysis was carried out per journal for all authors publishing at least one paper not deposited in ArXiv, and for authors with at least 5 non-deposited articles.

**Figure 4: Distribution for deposited versus non-deposited papers of authorships in *Physical Review B* among author citation impact quartiles**

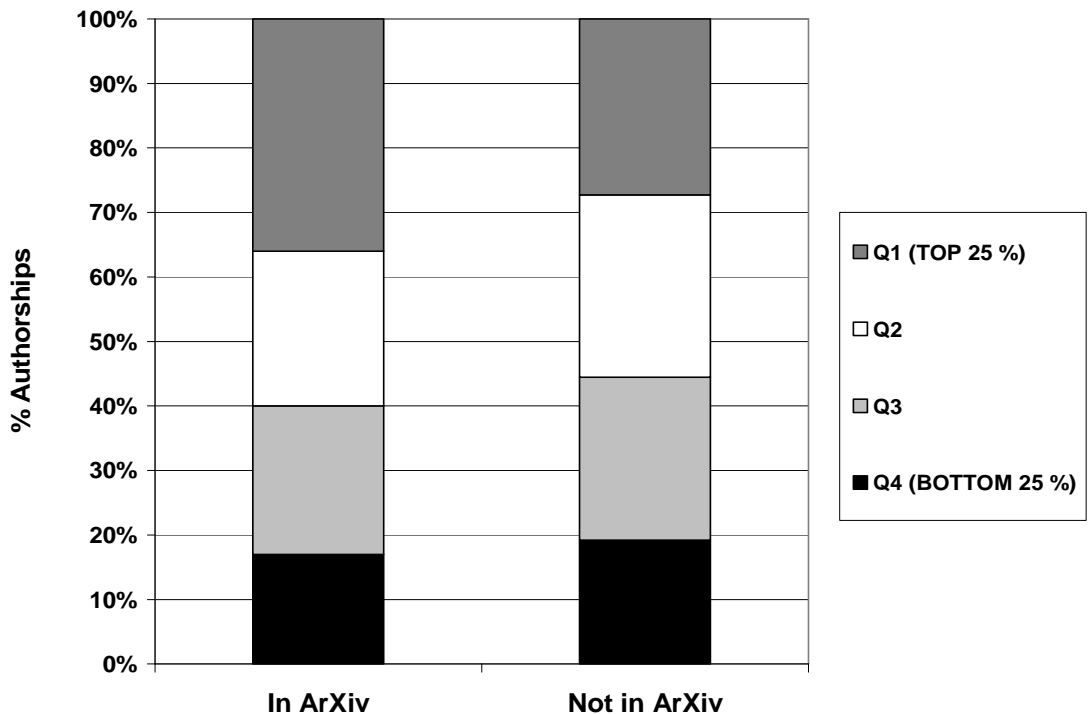

As an illustration, **Figure 4** presents the outcome of this analysis for the journal *Physical Review B*, and for authors publishing in that journal at least one paper that was not deposited in ArXiv. It shows that authors in the highest citation impact quartile ('top' authors) account for 36 per cent of authorships in the set of papers deposited in ArXiv-CM, and for 27 per cent in the set of non-deposited papers. The fact that both percentages are above 25 per cent reflects that top authors tend to publish more papers than less prominent authors do. For authors in the bottom 25 per cent of the citation impact distribution, the percentage of authorships in deposited and non-deposited papers are 17 and 19, respectively. Authors in this quartile tend to publish less papers than more prominent authors do. All journals listed in Table 1 show for both author productivity levels the same pattern: 'top' authors in terms of average citation impact per non-deposited paper are overrepresented in ArXiv (See Note 1).

In order to correct for this phenomenon, ArXiv Citation Impact Differentials (CID) were calculated at the level of an *individual* author publishing in a journal, and the *median* ArXiv CID was determined over publishing authors. This can be done for authors publishing at least one paper deposited in ArXiv-CM and at least one paper not deposited in that archive. **Figure 5** gives the results for *Physical Review B*.

**Figure 5: ArXiv Citation Impact Differentials over authors publishing in Physical Review B**

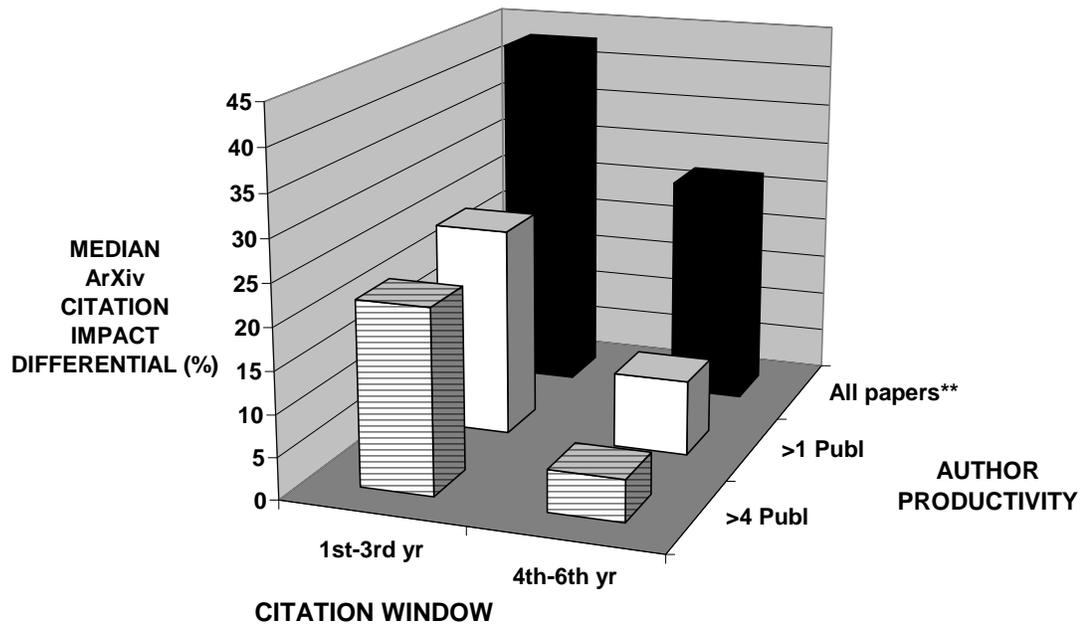

**Legend to Figure 5**: ** The percentages for All papers are not median values over authors, but the overall CID for the journal as a whole, presented in Table 1.

Figure 5 reveals the following patterns. Firstly, CID rates calculated for the fourth to sixth year after publication date tends to be lower than those based on the first to third year after publication. This phenomenon was already observed in Section 3.2, where it was ascribed to an early view effect. Secondly, compared to the CID rate for the journal as a whole, the median CID values over authors are lower, and decrease with increasing author productivity. For authors with more than 4 publications, applying the fourth to sixth year citation window, the median CID is 5 per cent. In this case there is hardly any 'impact advantage' of papers deposited in ArXiv.

**Table 2** presents the outcomes for each of the 6 journals with the largest number of papers deposited in ArXiv-CM. It shows a large variability among journals. It needs emphasising that the number of authors over which median CID values are calculated varies substantially among journals, citation windows and author productivity thresholds. It is rather low in several cases. But the general pattern is similar to that of *Physical Review B*: calculating median values over authors leads in most cases to a substantial reduction of the ArXiv Citation Impact Differentials compared to the overall rates presented in Table 1, and the more so if citations are counted during the fourth to sixth year after publication.

**Table 2:    Median ArXiv Citation Impact Differentials over authors**

| *Journal* | *1-3 yrs after publ date* | | | | *4-6 yrs after publ date* | | | |
|---|---|---|---|---|---|---|---|---|
| | *>1 publ* | | *>4 publ* | | *>1 publ* | | *>4 publ* | |
| | N | *CID* | *N* | *CID* | *N* | *CID* | *N* | *CID* |
| Physical Review B | 7,741 | *25 %* | 5,158 | *22 %* | 4,424 | *9 %* | 2,813 | *5 %* |
| European Physical Journal B | 394 | *67 %* | 74 | 29 % | 163 | *57 %* | 18 | 68 % |
| Journal of Physics – Cond. Matter | 874 | *40 %* | 332 | *37 %* | 441 | *29 %* | 146 | 29 % |
| Physica B – Condensed Matter | 766 | *6 %* | 406 | 11 % | 465 | *2 %* | 213 | 0 % |
| Solid State Communications | 445 | *40 %* | 153 | 29 % | 230 | *33 %* | 72 | 13 % |
| Internat Journal of Modern Physics B | 259 | *38 %* | 54 | *67 %* | 121 | *40 %* | 14 | -14 % |
| All 24 journals | 11,937 | *29 %* | 6,747 | *24 %* | 6,495 | *14 %* | 3,528 | *7 %* |

**Legend to Table 2**: N: Number of authors. Values in the cells are median values of the ArXiv Citation Impact Differential (CID) over authors publishing in a journal. Values printed in bold and italic are significantly different from 0 at p=0.01 according to the Sign Test. The second row indicates the publication thresholds for author productivity (>1 or >4 publications). The last row gives outcomes for all 24 journals aggregated. These median values are calculated on a journal-by journal-basis. In other words, CID was calculated for each author's publication oeuvre in a particular journal, and the median was calculated over all author–journal pairs in which the number of papers exceeded the publication threshold.

Comparing the median CID values over productive authors (i.e. authors with more than 4 papers) to the overall values for a journal as a whole, the average relative reduction rate over the 6 journals in Table 2 amounts to 56 per cent if citations are counted during the first three years after publication date, and 60 per cent if they are counted during the fourth to sixth year. *European Physical Journal B* obtained the lowest reduction rate (23 per cent) and *Physica B* the highest (93 per cent).

Figure 6 shows for the latter two journals and for *Physical Review B* the distribution of median CID among authors. The distribution of the first journal is almost symmetrical around the class with midpoint 0. *Physica B* has a large share of authors of which the papers deposited in ArXiv have a zero average impact, while their papers not deposited in that archive have impact above zero. These authors are included in the class with midpoint –200. For *European Physical Journal B* it is the other way around: there are relatively many authors for which the impact of their papers deposited in ArXiv is above zero, while that of their non-deposited articles is zero. These are included in the class with midpoint +200.

**Figure 6: Distribution of median CID values among authors for three journals**

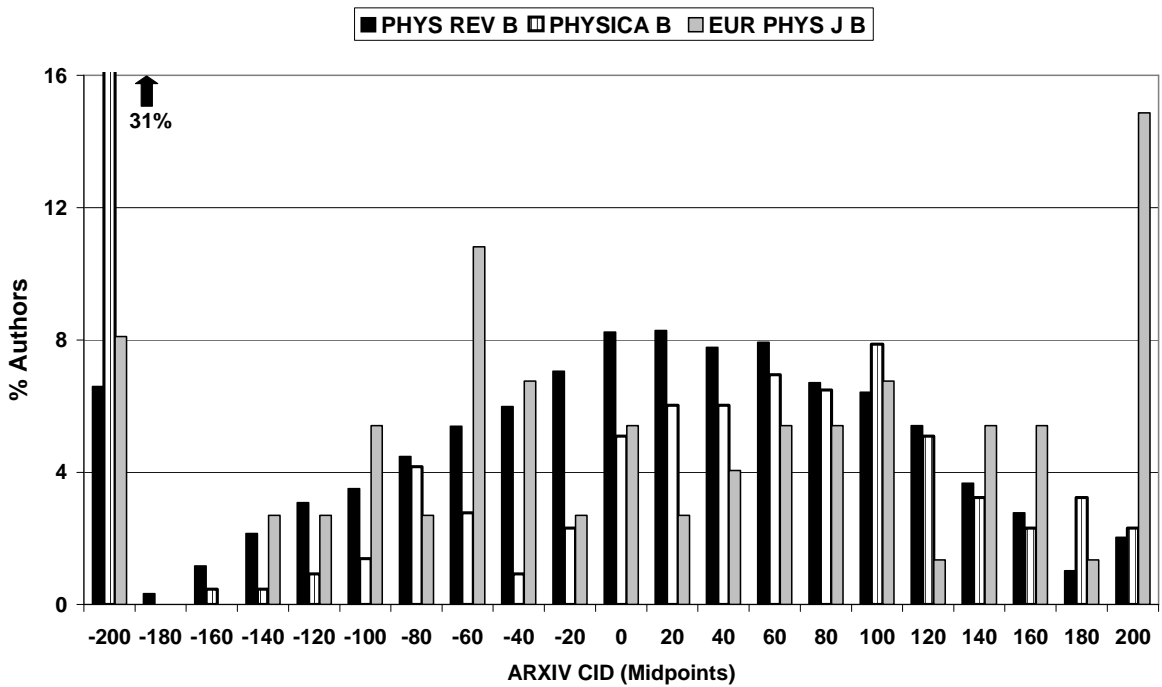

CID values for all authors with at least one paper deposited in ArXiv and one non-deposited paper are generally higher than those calculated for more productive authors, publishing more than 4 papers. In order to explain these differences, it must be noted that a substantial fraction of papers with at least one less productive author is co-authored by at least one productive author. For instance, for *Physical Review B*, 11 per cent has only productive authors, while 51 per cent has only less productive authors, and 38 per cent both a productive and a less productive one. The fraction of papers of less productive authors co-authored by a productive one is therefore 43 per cent. A first relevant finding is that this fraction is for papers deposited in ArXiv much higher than it is for non-deposited articles: 39 versus 82 per cent. All other 5 journals studied in detail in this paper showed a similar pattern.

A second finding was that the average citation impact of papers with only productive authors tends to be higher than that of articles with both productive and less productive ones, which on its turn is higher than the impact of papers authored by less productive authors only. This pattern was in most cases found for all 6 journals, for both citation windows, and both for papers deposited in ArXiv and for non-deposited papers. For instance, for papers published in Physical Review B, counting citations during the fourth to sixth year after publication, these three impact ratios amount to 15.1, 13.6 and 7.8, respectively for papers deposited in ArXiv, and 13.1, 12.3 and 8.7, respectively, for articles not deposited in that archive.

# 4 Discussion and conclusions

The analysis of ArXiv citation impact differentials (CID) for the collection of all 24 condensed matter physics journals presented in Figure 1 showed that the OA/non-OA Impact ratios as defined by Harnad and Brody (2004) fluctuated for papers between 1992 and 2003 around a level of 200 per cent and increased to 450 for papers published in 2005. These outcomes are in the same order of magnitude as those given by these two authors for 'All Physics Fields' and for 'Nuclear and Particle Physics', even though it is uncertain whether they included citations to ArXiv versions in their counts. Differences between the outcomes of the two studies may reflect differences among research fields.

The observation that ArXiv CID calculated for a set of papers varies with the age of those papers is crucial. The differences between the citation age distributions of deposited and non-deposited ArXiv-CM papers presented in Figure 2 can to a large extent – though not fully – be explained by the publication delay of about six months of non-deposited articles compared to papers deposited in ArXiv. This outcome provides evidence for an *early view effect* upon citation impact rates, and consequently upon ArXiv citation impact differentials. The early view effect is caused by the fact that colleagues in the field start the process of reading a paper, processing its information, and citing it in their own articles, *earlier* if a paper is deposited in ArXiv, because of its earlier availability.

The early view effect explains why CID values for recent years 2004 and 2005 are so much higher than those for earlier years. The observation that CID of journals calculated during the fourth to sixth year after publication are on average about 20 per cent lower than those calculated for the first three years after publication date, should also be attributed to an early view effect. The outcomes illustrate that a citation time window of 18 months, as applied in a recent article by Eysenbach (2006) may not be sufficiently long to adequately capture how 'OA versus non-OA impact ratios' vary with the age of cited articles.

Figure 4 provided evidence that prominent authors – measured per journal by the average citation impact of their papers *not* deposited in ArXiv –, are statistically overrepresented in the bylines of papers deposited in ArXiv. Therefore, it is appropriate to calculate CID rates at the level of individual authors. But such an analysis is to some extent hampered by the fact that the numbers of authors in the analysis on a journal-by-journal basis are in a number of cases rather low, especially if the publication productivity threshold is set to 4.

The calculation of *median* ArXiv Citation Impact Differentials *over authors* leads for the 6 journals presented in Table 1 on average to a reduction in CID with 56-60 per cent compared to the overall CID for a journal. This outcome suggests a strong quality bias in ArXiv Citation Impact Differentials or 'OA versus non-OA impact ratios'.

Considering more productive authors, and calculating citation impact during the fourth to sixth year after publication date, the median CID rates over authors in the 6 journals do not significantly differ from zero, except in the case of the 5 per cent rate for *Physical Review B* authors. For two journals CID is zero or even negative, for two other journals it is between 13 and 29 per cent, while for one journal it is 68 per cent. It needs emphasising that the two extreme values (–14 per cent for *International Journal of Modern Physics B* and + 68 per cent for *European Physical Journal B*) are based on low numbers of authors (18 and 14, respectively).

Median CID values for *all* authors publishing at least one paper deposited in ArXiv and one non-deposited paper were found to be higher than those for productive authors and are for all 6 journals significantly positive. This outcome can be explained by the finding that less productive authors' papers deposited in ArXiv are relatively more often co-authored with productive authors than their non-deposited papers, and that papers co-authored by productive *and* less productive authors – regardless of whether these papers were deposited in ArXiv or not – tend to have a higher citation impact than articles authored solely by less productive authors. Since senior authors tend to be more productive in terms of numbers of published papers, and tend to generate per paper a higher citation impact than junior authors, this effect can be interpreted as a *quality bias*: more productive, influential senior authors are overrepresented in the bylines of the papers deposited in ArXiv and authored by junior (or in any case less productive) researchers.

The conclusion is that, controlling for quality bias and early view effect, in the sample of 6 journals analysed in detail in this study, there is no sign of a general 'open access advantage' of papers deposited in ArXiv-CM. The empirical findings presented in this paper do provide evidence that ArXiv *accelerates* citation. This is actually a primary function of a pre-print archive, and the outcomes reveal that ArXiv is successful in carrying out this function. Accelerating communication is definitely a positive effect of ArXiv. But the findings presented in this study suggest that this acceleration is due to the fact that that ArXiv makes papers *earlier* available rather than that it makes papers *freely* available.

It would be illuminating to further analyse early view effects also in other publication environments. For instance, in electronic journal collections of publishers adopting a subscription-based business model, journals issues may be available electronically several months before their formal publication. The time interval between electronic and formal publication date varies across journals, and changes over time. Selecting appropriate groups of journals and control groups, the early view effect could be further studied and quantified.

The effect of delays in the publication process and their effect upon age distributions of citations is an important topic of bibliometric and informetric research (e.g., Egghe and Rousseau (2000)). A further analysis and modelling of such phenomena, particularly in relation to preprint archiving, falls beyond the scope of this article and awaits further research. This research should also include citations within ArXiv, i.e., citations from one ArXiv paper to another. These data were not available in the study presented in this paper.

The analysis of quality bias focused on the extent to which prominent authors are overrepresented in the bylines of papers deposited in ArXiv. A second dimension, the extent to which authors – be it prominent or less so – tend to deposit their better papers in ArXiv had not been examined and awaits further research. This can only be done by using measures of author prominence that are not based upon citation impact.

## Acknowledgements

The main lines of the work described in this paper were presented at the First International Conference of the Association of Publishers in Europe (APE), Berlin (Germany), 4-5 April, 2006, and at the Open Scholarship 2006 Conference in Glasgow (UK), 18-20 October, 2006. The author is grateful to numerous participants at these conferences for their comments, and to his CWTS colleague Martijn Visser for his technical assistance and stimulating discussions. The research described in this paper is partly funded by Elsevier within the framework of a research project 'Reference behaviour of scientific authors', carried out at CWTS. Several techniques applied in this paper were developed in a research project funded by the Netherlands Organisation for Scientific Research (NWO) on the development of bibliometric indicators for the assessment of research performance in the field computer science. The author of this paper emphasises that both granting organizations respect the academic and entirely independent position of the researchers involved in these projects.

## Notes

***Note 1***. Authors publishing only papers that were deposited in ArXiv are not included in this analysis. The percentage of these authors strongly depends upon the publication productivity threshold applied. Considering all authors with at least one paper in a journal, it was 23 per cent for *European Physical Journal B*, 6 per cent for *Physical Review B* and less than 4 per cent for the other 4 journals listed in Table 1. For authors with >4 publications it was 18 per cent for *European Physical Journal B*, 1.2 per cent for Physical Review B, and almost zero for the other five journals.